\documentclass[final,onecolumn,sort,compress]{elsarticle}
\usepackage{epstopdf}
\usepackage{cmap}
\usepackage[english]{babel}
\usepackage{amsmath,amssymb,amsfonts}
\usepackage[colorlinks=true,unicode,pdfhighlight=/P]{hyperref}

\begin{document}
\begin{frontmatter}
\title{Analysis of stationary points and their bifurcations in the ABC flow}
\author{A. A. Didov}
\address{School of Natural Sciences, Far Eastern Federal University, 690090 Vladivostok, Russia}%
\address{Laboratory of Nonlinear Dynamical Systems, Pacific Oceanological Institute of the Russian Academy of Sciences, 690041 Vladivostok, Russia}
\ead{Kedr\_ad@mail.ru}

\author{M. Yu. Uleysky}
\address{Laboratory of Nonlinear Dynamical Systems, Pacific Oceanological Institute of the Russian Academy of Sciences, 690041 Vladivostok, Russia}
\ead{uleysky@poi.dvo.ru}

\begin{abstract}
Analytical expressions for coordinates of stationary points and conditions for their
existence in the ABC flow are received. The type of the stationary points is shown analytically to be saddle-node. Exact expressions for eigenvalues and eigenvectors of the stability matrix
are given. Behavior of the stationary points  along the bifurcation lines is described.
\end{abstract}

\end{frontmatter}

\section{Introduction}
The well-known Arnold--Beltrami--Childress (ABC) flow is a steady solution of the Euler equations
for a steady incompressible flow of Newtonian fluids. Furthermore, the ABC flow can be
considered  as a solution of the Navier--Stokes equations if an external body force
$f=-\nu \Delta\vec{V}=\nu\vec{V}$, just compensating viscous losses, is applied.
Arnold \cite{Arnold1965} firstly suggested chaos in the field lines
and, therefore, in trajectories in that three-dimensional steady flow. The chaotic aspect then
has been developed and advanced for a wide class of two-dimensional unsteady flows
under the name ``chaotic advection''
\cite{Aref,Meleshko1996,Aref2002,KP06,Chaos07,Koshel08,Budyansky2009,PRE10}.

The ABC flow has been studied by many authors. Dombre et al.~\cite{Dombre1986} investigated the
ABC flow both analytically and numerically for different values of the real control parameters
$A$, $B$, $C$.
Henon \cite{Henon1966} provided a numerical evidence for chaos in the special case
at $A=\sqrt{3}$, $B=\sqrt{2}$ and $C=1$. Independently Childress~\cite{Childress1970} considered
the special case as a model for the kinematic dynamo effect where $A = B = C = 1$
(this version is useful in dynamo theory).
In 1993, Zhao et al.~\cite{Zhao1993} obtained an analytical criteria for existence of chaotic
and resonant streamlines in the ABC flow by the Melnikov' method \cite{Melnikov1963}.
In 1998, Huang et al.~\cite{Huang1998} obtained an explicit analytical criterion for existence
of chaotic streamlines in the ABC flow.
Ziglin et al.~\cite{Ziglin2003} proved that the ABC flow at $A = B = C$
has no real-analytic first integral.
Later Maciejewski~\cite{Maciejewski2002} has investigated non-integrability of the
ABC flow at the condition
$A^2=B^2$ for $ABC\neq0$, and it was proved that the set does not possess the first real
meromorphic integral for $A^2/B^2<2$. The integrable case (when one of the parameters $A$, $B$ or $C$ vanishes)
has been considered in Ref.~\cite{Dombre1986}.
Another case of the absence and existence of the first integrals
has been considered in Refs.~\cite{Llibre2012, Ziglin1996, Ziglin1998}.
Brummell~\cite{Brummell2001} has investigated some properties of the time-dependent ABC flow.
In Ref.~\cite{DidovChaos2018}  obtained an explicit analytical criterion and
numerical evidences for the existence of the $n : m$ resonances in the ABC flow.

Arnold and Korkina~\cite{Arnold1983}, Galloway and Frisch~\cite{Galloway1984}, and
Moffatt and Proctor~\cite{Moffatt1985} performed numerical and analytical studies of
the dynamo action at finite values of the conductivity.
They have shown that the ABC flow can excite a magnetic field.
However, it was shown that the generated magnetic field looks as a combination
of cigar-like structures and can hardly be considered as a large-scale one.
The Arnold--Beltrami--Childress flow is a prototype for the fast dynamo action, essential
to the origin of magnetic field for large astrophysical objects, and dynamo properties can be
investigated by varying the magnetic Reynolds number $R_m$.
Recently, some properties the ABC flow have been investigated by Galloway~\cite{Galloway2012}.
Bouya~\cite{Bouya2013} has investigated properties of the ABC flow for the extended region from
$R_m < 1600$ to $R_m < 25 000$. In 2013, Jones and Gilbert~\cite{Jones2013}
have studied in detail symmetries of the various dynamo branches up
to $R_m = 10^4$. In 2015, Bouya~\cite{Bouya2015} has investigated the kinematic dynamo action
up to $R_m = 5\cdot 10^5$.  
Solution for stationary points in the special case has been obtained in Ref.~\cite{Ershkov2016}. It has been proved the existence of 1 point for two partial cases of  parameters $A$, $B$, $C$: 1) $A=B=1$; 2) $C=1$ $(A^2+B^2=1)$. 

In this paper, we find stationary points in the general case and criteria for their existence.
The behavior of streamlines in vicinity of stationary points will be discussed as well.
Exact analytic expression for eigenvalues and eigenvectors of the stability matrix will
be given for the first time.

\section{Analytical solution of the ABC flow}
Let us write the autonomous set of differential equations for the ABC flow:
\begin{equation}
\label{1}
\begin{aligned}
\frac{dx}{dt}=V_x = A\sin(z)+C\cos(y),\\
\frac{dy}{dt}=V_y = B\sin(x)+A\cos(z),\\
\frac{dz}{dt}=V_z = C\sin(y)+B\cos(x),
\end{aligned}
\end{equation}
where $A$, $B$ and $C$ are real parameters.
This flow is periodic on all three variables with period $2\pi$, so, we consider only one cubic box
$x$, $y$, $z\in[-\pi,\pi)$.
Let $A$, $B$ and $C$ are greater than or equal
to zero. If any of $A$, $B$ and $C$ parameter is negative,
we can translate the corresponding variable by $\pi$.
By using some results from~\cite{Dombre1986} we can assume
\begin{equation}
\label{Cond}
\begin{aligned}
1=A\geqslant B\geqslant C\geqslant0,
\end{aligned}
\end{equation}
then (\ref{1}) can be rewritten as
\begin{equation}
\label{2}
\begin{aligned}
&\frac{dx}{dt}=V_x = \sin(z)+C\cos(y),\\
&\frac{dy}{dt}=V_y = B\sin(x)+\cos(z),\\
&\frac{dz}{dt}=V_z = C\sin(y)+B\cos(x).
\end{aligned}
\end{equation}

Let us find the stationary points of the ABC flow system. Assuming $dx/dt=dy/dt=dz/dt=0$,
we get the set of algebraic equations
\begin{equation}
\label{3}
\begin{aligned}
&\sin(z_0)=-C\cos(y_0),\\
&B\sin(x_0)=-\cos(z_0),\\
&C\sin(y_0)=-B\cos(x_0),
\end{aligned}
\end{equation}
where $M(x_0, y_0, z_0)$ is a stationary point.
Squaring (\ref{3}) and using identity $\cos^2(\alpha)+\sin^2(\alpha)=1$, we get
\begin{equation}
\label{4}
\begin{aligned}
&\sin^2(z_0)=C^2(1-\sin^2(y_0)),\\
&B^2\sin^2(x_0)=1-\sin^2(z_0),\\
&C^2\sin^2(y_0)=B^2(1-\sin^2(x_0)).
\end{aligned}
\end{equation}
After replacement
\begin{equation}
\label{5}
X=\sin^2(x_0),\quad
Y=\sin^2(y_0),\quad
Z=\sin^2(z_0),
\end{equation}

we obtain the solution of (\ref{4}) is the following form:
\begin{equation}
\label{7}
X=\frac{B^2-C^2+1}{2B^2},\quad
Y=\frac{C^2+B^2-1}{2C^2},\quad
Z=\frac{C^2-B^2+1}{2}.
\end{equation}
\begin{figure}[!htb]
\begin{center}
\includegraphics[width=1\textwidth]{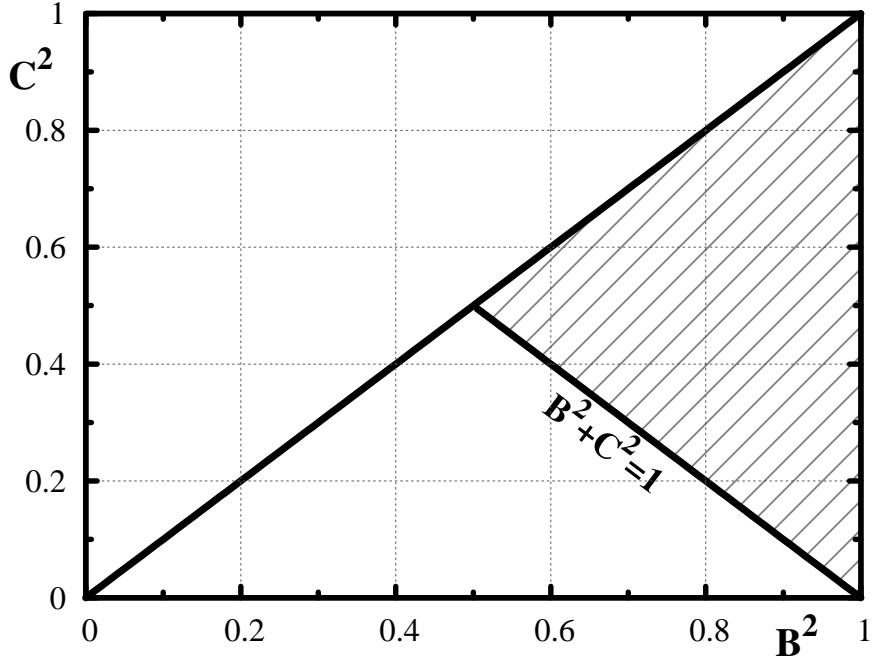}
\end{center}
\caption{The parametric space of the set~(\ref{3}) with conditions~(\ref{Cond}) (region under line $C^2=B^2$). Hatched area
is zone of existence of stationary points.}
\label{par1}
\end{figure}
Substitute $X$, $Y$ and $Z$ from (\ref{7}) to (\ref{5}), we get
\begin{equation}
\label{8}
\begin{aligned}
x_0 = \delta_x \arcsin\left(P_x \right),
\qquad
x_0 = \delta_x \left\{\pi-\arcsin\left(P_x \right)\right\},\\
y_0 = \delta_y \arcsin\left(P_y \right),
\qquad
y_0 = \delta_y \left\{\pi-\arcsin\left(P_y \right)\right\},\\
z_0 = \delta_z \arcsin\left(P_z\right),
\qquad
z_0 = \delta_z \left\{\pi-\arcsin\left(P_z\right)\right\},
\end{aligned}
\end{equation}
where
\begin{equation}
\label{9}
\begin{aligned}
P_x=\sqrt{\frac{B^2-C^2+1}{2B^2}},
\qquad
\delta_x=\pm1,\\
\qquad
P_y=\sqrt{\frac{B^2+C^2-1}{2C^2}},
\qquad
\delta_y=\pm1,\\
P_z=\sqrt{\frac{-B^2+C^2+1}{2}},
\qquad
\delta_z=\pm1.
\end{aligned}
\end{equation}
To allow existence of solution of the set (\ref{3}) the parameters $P_x$, $P_y$ and $P_z$
must be between zero and one.
This is true if
\begin{equation}
\label{10}
C^2\geqslant 1-B^2
\end{equation}
in additional to inequality~(\ref{Cond}).
Region of existence of solution in the parametric space $B$, $C$ by the hatched area on Fig.~\ref{par1}.

Let us write (\ref{8}) in the general form
\begin{equation}
\label{11}
\begin{aligned}
x_0 = \delta_x \left[\frac{1}{2}\left(1-\gamma_x\right)\pi+\gamma_x\arcsin\left(P_x \right)\right],\\
y_0 = \delta_y \left[\frac{1}{2}\left(1-\gamma_y\right)\pi+\gamma_y\arcsin\left(P_y \right)\right],\\
z_0 = \delta_z \left[\frac{1}{2}\left(1-\gamma_z\right)\pi+\gamma_z\arcsin\left(P_z\right)\right].
\end{aligned}
\end{equation}
The coefficients $\delta_i$ and $\gamma_i$ ($i=x, y, z$) are determined by the
following rule:
\begin{equation}
\label{12}
\delta_i =
\begin{cases}
+1, & i \in[0,\pi), \\
-1, & i \in[-\pi, 0),
\end{cases}\qquad
\gamma_i =
\begin{cases}
+1, & i \in[-\pi/2,\pi/2), \\
-1, & i \in[-\pi,-\pi/2)\cup[\pi/2,\pi).
\end{cases}
\end{equation}

There are 64 combinations of the coefficients $\delta_i$ and $\gamma_i$, but only 8
satisfy the initial equations (\ref{3}), and they are presented in Table \ref{tab1}. Each point has its own unique
combination of the coefficients $\delta_i$, so each octant of the phase space box contains exactly one stationary
point.
\begin{table}[!ht]
\caption{The sign of $\delta_i$ and $\gamma_i$ for
different stationary points.}
\label{tab1}
\begin{center}
\begin{tabular}{|c|c|c|c|c|c|c|c|c|}
\hline
Solution & 1 & 2 & 3 & 4 & 5 & 6 & 7 & 8 \\
\hline
\hline
Parameter & \multicolumn{8}{c|}{Sign} \\
\hline
$\delta_x$ & $-$ & $-$ & $+$ & $+$ & $+$ & $+$ & $-$ & $-$ \\
\hline
$\gamma_x$ & $-$ & $+$ & $+$ & $-$ & $-$ & $+$ & $+$ & $-$ \\
\hline
\hline
$\delta_y$ & $+$ & $-$ & $-$ & $+$ & $+$ & $-$ & $-$ & + \\
\hline
$\gamma_y$  & $-$ & $-$ & $-$ & $-$ & $+$ & $+$ & $+$ & $+$ \\
\hline
\hline
$\delta_z$ & $+$ & $+$ & $+$ & $+$ & $-$ & $-$ & $-$ & $-$ \\
\hline
$\gamma_z$ & $+$ & $+$ & $-$ & $-$ & $-$ & $-$ & $+$ & $+$ \\
\hline
\end{tabular}
\end{center}
\end{table}

Now we briefly consider behavior of the stationary points~(\ref{11})
on the bifurcation line (\ref{10}) separating region of existence of stationary points
in the parametric space. It follows from~(\ref{11}) the following stationary points will merge:
(1,~2), (3,~4), (5, 6) and (7, 8).
So, (\ref{2}) have 4 stationary points on the bifurcation line~(\ref{10}).

The set (\ref{2}) is integrable at the point ($B=1$, $C=0$). In this case the solutions (1, 2, 3, 4)
and (5, 6, 7, 8) are merge and form stationary lines
($x=\pi/2$, any $y$, $z=-\pi$) and ($x=-\pi/2$, any $y$, $z=0$).

\section{Stability analysis of solutions of the ABC flow}

Now we consider behavior of phase trajectories near the stationary points~(\ref{11}).
Linearizing the set (\ref{2}) in vicinity of the stationary points (\ref{11}), we obtain equations
\begin{equation}
\label{14}
\begin{aligned}
\dot{\Delta x} = &\cos\left(\delta_z \left[\frac{1}{2}\left(1-\gamma_z\right)\pi+\gamma_z\arcsin P_z\right]
\right)\Delta z-\\
&\delta_yC\sin\left( \frac{1}{2}\left(1-\gamma_y\right)\pi+\gamma_y\arcsin P_y
\right)\Delta y,\\
\dot{\Delta y} = &B\cos\left(\delta_x \left[\frac{1}{2}\left(1-\gamma_x\right)\pi+\gamma_x\arcsin P_x
\right]\right)\Delta x-\\
&\delta_z\sin\left( \frac{1}{2}\left(1-\gamma_z\right)\pi+\gamma_z
\arcsin P_z\right)\Delta z,\\
\dot{\Delta z} = &C\cos\left(\delta_y \left[\frac{1}{2}\left(1-\gamma_y\right)\pi+\gamma_y\arcsin P_y
\right]\right)\Delta y-\\
&\delta_xB\sin\left( \frac{1}{2}\left(1-\gamma_x\right)\pi+\gamma_x
\arcsin P_x \right)\Delta x,
\end{aligned}
\end{equation}
where $\Delta x=x-x_0$, $\Delta y=y-y_0$ and $\Delta z=z-z_0$.
We can omit in (\ref{14}) the coefficient $\delta_i$ in the arguments of cosines.
Using the trigonometric identities
$\sin(\pi-\alpha)=\sin(\alpha)$, $\cos(\pi-\alpha)=-\cos(\alpha)$ and expressing cosines by sines,
we get
\begin{equation}
\label{15}
\begin{aligned}
&\dot{\Delta x} = \gamma_z\sqrt{1-P_z^2}\Delta z-\delta_yC P_y\Delta y,\\
&\dot{\Delta y} = \gamma_xB\sqrt{1-P_x^2}\Delta x-\delta_zP_z\Delta z,\\
&\dot{\Delta z} = \gamma_yC\sqrt{1-P_y^2}\Delta y-\delta_xBP_x\Delta x.
\end{aligned}
\end{equation}

The solution of equations (\ref{15}) is determined by the roots of characteristic equation
\begin{multline}
\label{16}
\begin{vmatrix}
-\lambda          & -\delta_yC P_y &               \gamma_z\sqrt{1-P_z^2} \\
\gamma_xB\sqrt{1-P_x^2} &         -\lambda  &          -\delta_zP_z \\
-\delta_xBP_x &       \gamma_yC\sqrt{1-P_y^2} &           -\lambda
\end{vmatrix} =\\
\lambda^3+\lambda\left(\delta_x\gamma_zBP_x\sqrt{1-P_z^2}+
\delta_z\gamma_y CP_z\sqrt{1-P_y^2}+\delta_y\gamma_x BC P_y
\sqrt{1-P_x^2}\right)-\\
-BC\left(\gamma_x\gamma_y\gamma_z\sqrt{1-P_x^2}\sqrt{1-P_y^2}\sqrt{1-P_z^2}
-\delta_x\delta_y\delta_zP_xP_yP_z\right)=0.
\end{multline}
Numerical values of the coefficients $\gamma_i$ and $\delta_i$ given in Table \ref{tab1}
and their multiplication are presented in Table \ref{tab2}. Let
\begin{multline}
\label{17}
\begin{aligned}
Q=&BP_x\sqrt{1-P_z^2}+C P_z\sqrt{1-P_y^2}+ BC P_y\sqrt{1-P_x^2}= \frac{1}{2}\left(1+C^2+B^2\right),\\
W=&BC\left(\sqrt{1-P_x^2} \sqrt{1-P_y^2}\sqrt{1-P_z^2}+P_xP_yP_z\right)=
\end{aligned}\\
=\frac{1}{\sqrt{2}}\sqrt{\left(B^2+C^2-1\right)\left(1-\left(B^2-C^2\right)^2\right)}.
\end{multline}
Finally, the characteristic equation (\ref{16}) has the following form:
\begin{equation}
\label{18}
\begin{aligned}
\lambda^3-\lambda Q - \xi W=0,
\end{aligned}
\end{equation}
where $\xi=\gamma_x\gamma_y\gamma_z=-\delta_x\delta_y\delta_z$ depends
on choice of solution of (\ref{3})
\begin{equation}
\label{19}
\xi =
\begin{cases}
+1, & \text{for solutions 1, 3, 5, 7,}\\
-1, & \text{for solutions 2, 4, 6, 8.}
\end{cases}
\end{equation}

\begin{table}[!ht]
\caption{The sign of multiplication of coefficients $\gamma_i$
and $\delta_i$.}
\label{tab2}
\begin{center}
\begin{tabular}{|c|c|c|c|c|c|c|c|c|}
\hline
Solution& 1 & 2 & 3 & 4 & 5 & 6 & 7 & 8 \\
\hline
\hline
Parameter & \multicolumn{8}{c|}{Sign} \\
\hline
\hline
$\delta_x\gamma_z$ & $-$ & $-$ & $-$ & $-$ & $-$ & $-$ & $-$ & $-$ \\
\hline
$\delta_z\gamma_y$ & $-$ & $-$ & $-$ & $-$ & $-$ & $-$ & $-$ & $-$ \\
\hline
$\delta_y\gamma_x$ & $-$ & $-$ & $-$ & $-$ & $-$ & $-$ & $-$ & $-$ \\
\hline
\hline
$\delta_x\delta_y\delta_z$ & $-$ & $+$ & $-$ & $+$ & $-$ & $+$ & $-$ & $+$ \\
\hline
$\gamma_x\gamma_y\gamma_z$ & $+$ & $-$ & $+$ & $-$ & $+$ & $-$ & $+$ & $-$ \\
\hline
\end{tabular}
\end{center}
\end{table}

The equation (\ref{18}) has two variants (for cases $\xi=-1$ and $\xi=1$).
We consider each case individually. Let $\xi=-1$, then we get polynomial (\ref{18}) in the form
\begin{equation}
\label{20}
\begin{aligned}
Y_1(\lambda)=\lambda^3-\lambda Q + W=0.
\end{aligned}
\end{equation}
The polynomial (\ref{20}) has extreme points $\lambda_1=-\sqrt{\frac{Q}{3}}$
and $\lambda_2=\sqrt{\frac{Q}{3}}$.
Since $Y_1(0)>0$, then the polynomial (\ref{20}) has one real negative root. The two other
roots can be either complex conjugate pair with positive real part
($Y_1(\lambda_2)>0$) or real positive ones ($Y_1(\lambda_2)\leqslant0$).
We will prove that $Y_1(\lambda_2)\leqslant0$ for all values of parameters $B$ and $C$.
\begin{equation}
\label{21}
\begin{aligned}
Y_1(\lambda_2)=\frac{Q\sqrt{Q}}{3\sqrt{3}}-Q\sqrt{\frac{Q}{3}} + W \leqslant 0.
\end{aligned}
\end{equation}
After simplification and substitution (\ref{17}) to (\ref{21}), we get
\begin{equation}
\label{22}
\begin{aligned}
\frac{1}{3\sqrt{3}}\left(1+B^2+C^2\right)\sqrt{\left(1+B^2+C^2\right)}\geqslant\sqrt{\left(B^2+C^2-1
\right)\left(1-\left(B^2-C^2\right)^2\right)}.
\end{aligned}
\end{equation}
By squaring inequality (\ref{22}) we obtain
\begin{equation}
\label{23}
\begin{aligned}
\frac{1}{27}\left(1+B^2+C^2\right)^3\geqslant\left(B^2+C^2-1\right)\left(1-
\left(B^2-C^2\right)^2\right).
\end{aligned}
\end{equation}

Let
\begin{equation}
\label{24}
q=B^2+C^2,\qquad
p=B^2-C^2.
\end{equation}
In the region of existence of the solution parameters we have $q\in[1,2]$, $p\in[0,1]$.
Inequality~(\ref{23}) can be rewritten as follows:
\begin{equation}
\label{25}
\begin{aligned}
\frac{1}{27}\frac{\left(1+q\right)^3}{\left(q-1\right)}\geqslant1-p^2,
\end{aligned}
\end{equation}
Since $1-p^2\leqslant1$, we can strengthen the inequality (\ref{25})
\begin{equation}
\label{26}
\begin{aligned}
T(q)=\left(1+q\right)^3-27{\left(q-1\right)}\geqslant0.
\end{aligned}
\end{equation}
The inequality~(\ref{26}) is true, because $T(1)>0$ and $T(q_\text{min})=0$,
where $q_\text{min}=2$ is a minimum point of $T(q)$.
So equation (\ref{20}) has 3 real roots (two are positive and one is negative).
In the case $q=2$ and $p=0$ ($B=C=1$) equation (\ref{20}) has a multiple positive root.

\begin{figure}[!ht]
\begin{minipage}[h]{0.45\linewidth}
\center{\includegraphics[width=0.8\linewidth]{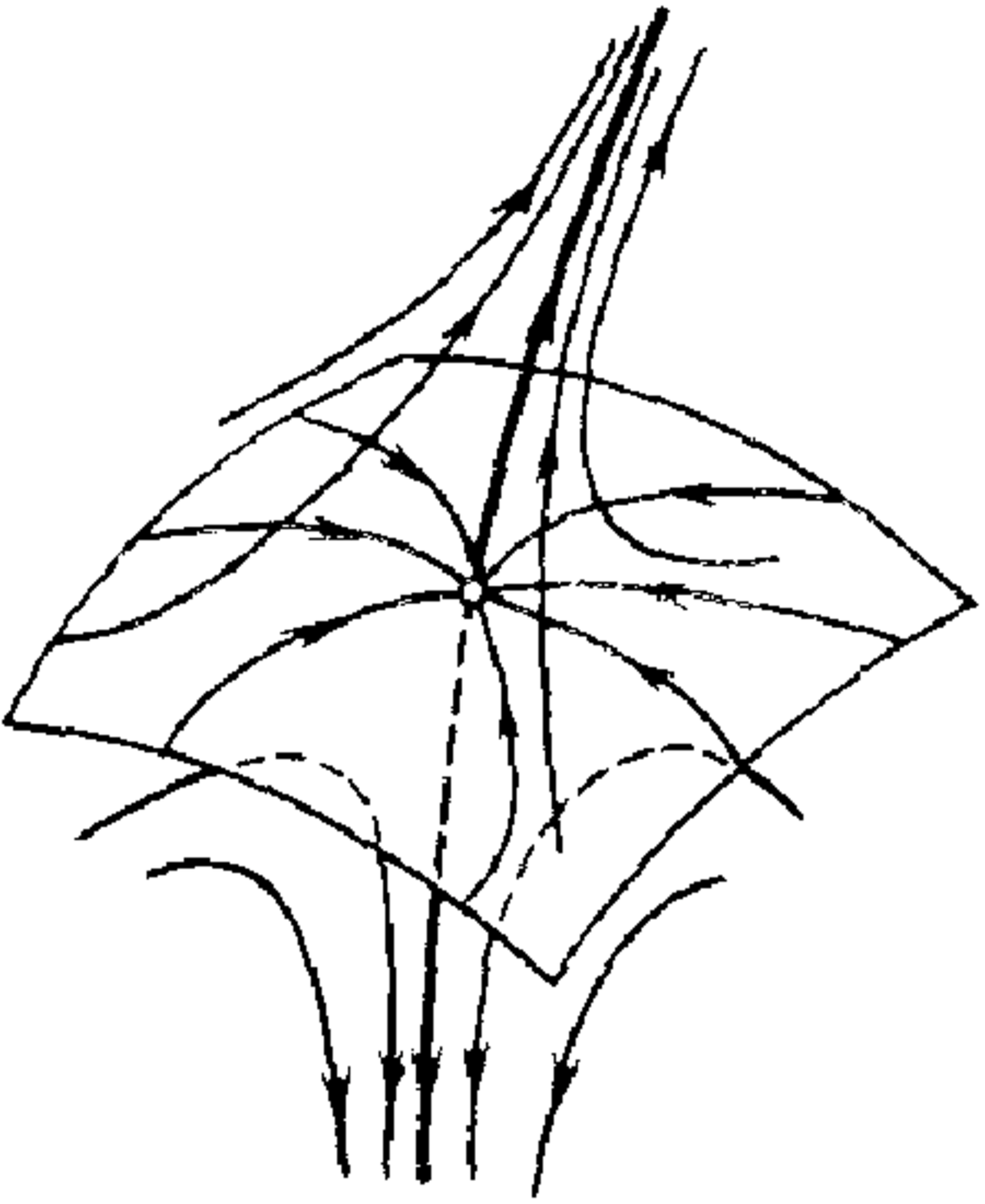}}
\caption{Streamlines in a vicinity of the stationary point with two negative eigenvalues
and a positive one.}
\label{image1}
\end{minipage}
\hfill
\begin{minipage}[h]{0.45\linewidth}
\center{\includegraphics[width=0.8\linewidth]{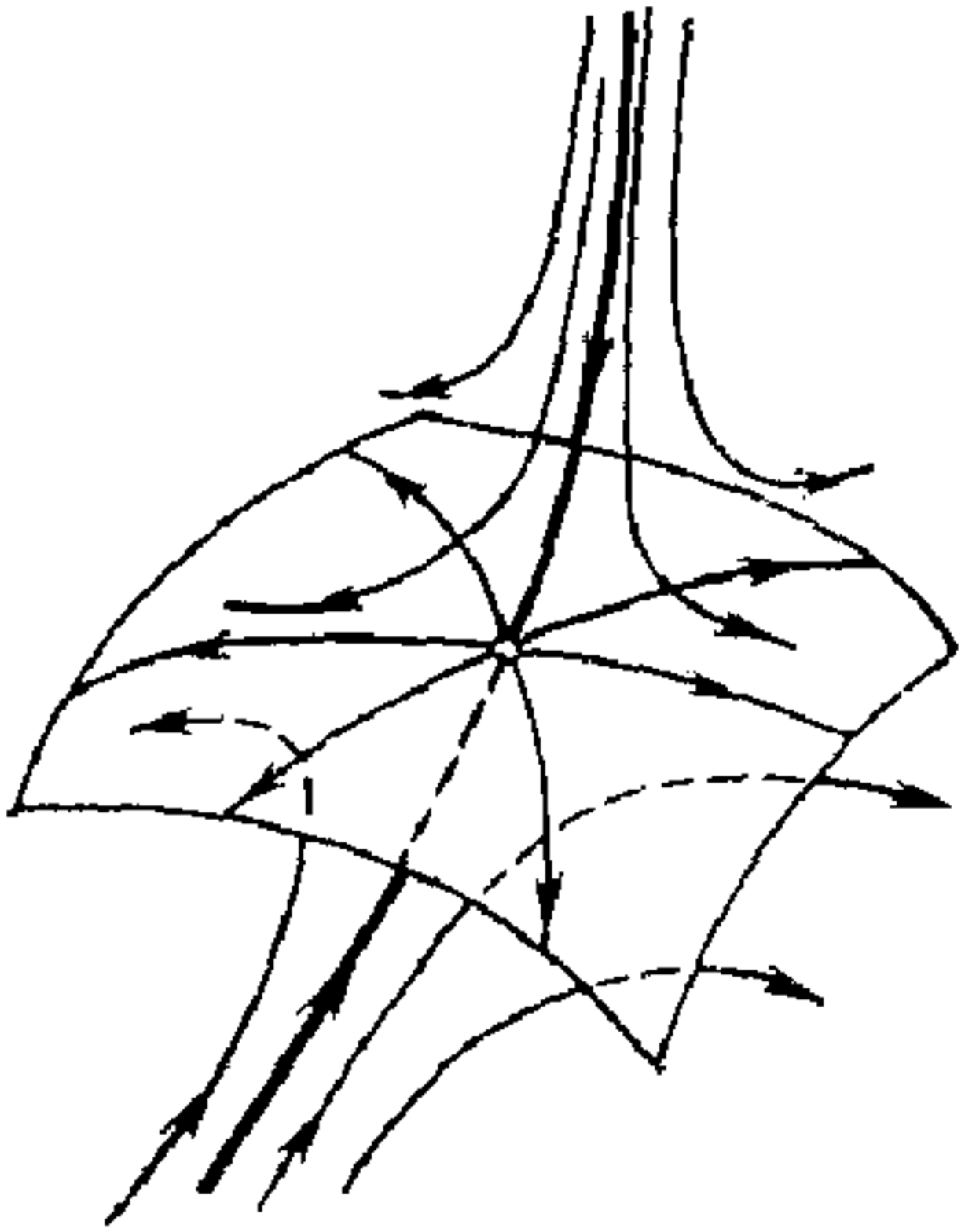}}
\caption{Streamlines in a vicinity of the stationary point with one negative eigenvalue
and two positive ones.}
\label{image2}
\end{minipage}
\end{figure}

Let us consider the case $\xi=+1$. The polynomial (\ref{18}) can be written as
\begin{equation}
\label{27}
\begin{aligned}
Y_2(\lambda)=\lambda^3-\lambda Q - W = Y_1(-\lambda) =  0.
\end{aligned}
\end{equation}
Because of $Y_2(\lambda)=Y_1(-\lambda)$, the polynomial (\ref{27}) has two negative real roots
and one positive real root.
The streamlines around the stationary points are shown in Figs.~\ref{image1} and ~\ref{image2}.

The equation for eigenvalues (\ref{28}) on bifurcation lines reduces
to $\lambda^3+\alpha\lambda=0$, so, one
of the eigenvalues is zero. All these stationary points are plane saddles.

\section{Eigenvalues}
In this section we obtain an exact solution for eigenvalues of the stability matrix.
We will use the Cardano method for the cubic equation expressed as
\begin{equation}
\label{28}
\begin{aligned}
\lambda^3+\alpha\lambda+\beta=0,
\end{aligned}
\end{equation}
where $\alpha=-Q$, $\beta=-\xi W$. Solution of (\ref{28}) is defined by the discriminant
\begin{equation}
\label{29}
\begin{aligned}
S = \frac{\beta^2}{4}+\frac{\alpha^3}{27}=\frac{216\left(q-1\right)\left(1-p^2\right)-8\left(1+q\right)^3}{1728}.
\end{aligned}
\end{equation}
The discriminant $S$ is real and the type of roots of Eq.~(\ref{28}) can be defined if we known the sign of $S$.
Now we show that $S$ is always less or equal to zero.
Let's consider the polynomial
\begin{equation}
\label{30}
\begin{aligned}
\mathcal{I}(q)=216\left(q-1\right)\left(1-p^2\right)-8\left(1+q\right)^3.
\end{aligned}
\end{equation}
On the edges of the interval $q\in[1,2]$ we have $\mathcal{I}(q)\leqslant0$.
Consequently $\mathcal{I}(q)$ is less or equal to zero for all $q\in[1,2]$
if $\mathcal{I}(q_1)\leqslant 0$, where $q_1=-1+3\sqrt{1-p^2}$ is a maximum point
of $\mathcal{I}(q)$. Let us consider
\begin{multline}
\label{31}
\mathcal{I}(q_1)=216\left(3\sqrt{1-p^2}-2\right)
\left(1-p^2\right)-8\left(3\sqrt{1-p^2}\right)^3=\\
=\left(\sqrt{1-p^2}-1\right)\left(1-p^2\right).
\end{multline}
Because of $\sqrt{1-p^2}\leqslant1$, we obtain that $\mathcal{I}(q)\leqslant0$
for all $q\in[1,2]$.

In the case $S\leqslant0$ the equation (\ref{28}) has three real roots
which can be found by the following way:
\begin{equation}
\label{32}
\begin{gathered}
\lambda_1=2\sqrt{\frac{-\alpha}{3}} \cos\left(\frac{F}{3}\right),\\
\lambda_2=2\sqrt{\frac{-\alpha}{3}} \cos\left(\frac{F}{3}+\frac{2\pi}{3}\right),\\
\lambda_3=2\sqrt{\frac{-\alpha}{3}} \cos\left(\frac{F}{3}+\frac{4\pi}{3}\right),\\
\end{gathered}
\end{equation}
where $F$ are defined by
\begin{equation}
\label{33}
F=
\begin{cases}
\arctan\left(\frac{2\sqrt{\vert S\vert}}{-\beta}\right), & \beta<0,\\
\arctan\left(\frac{2\sqrt{\vert S\vert}}{-\beta}\right)+\pi, & \beta\geqslant0.
\end{cases}
\end{equation}
All roots and $F$ are defined via the parameters $B$ and $C$ as
\begin{equation}
\label{34}
\begin{gathered}
\lambda_1=2\sqrt{\frac{\left(1+C^2+B^2\right)}{6}} \cos\left(\frac{F}{3}\right),\\
\lambda_2=2\sqrt{\frac{\left(1+C^2+B^2\right)}{6}} \cos\left(\frac{F}{3}+\frac{2\pi}{3}\right),\\
\lambda_3=2\sqrt{\frac{\left(1+C^2+B^2\right)}{6}} \cos\left(\frac{F}{3}+\frac{4\pi}{3}\right),\\
\end{gathered}
\end{equation}
\begin{equation}
\label{35}
F =
\begin{cases}
\arctan\left(2\sqrt{2}\sqrt{ \frac{\vert 216\left(B^2+C^2-1\right)\left(1-(B^2-C^2)^2\right)
-8\left(1+B^2+C^2\right)^3\vert}{1728\left(B^2+C^2-1\right)
\left(1-\left(B^2-C^2\right)^2\right)}}\right),& \xi = 1, \\
\pi-\arctan\left(2\sqrt{2}\sqrt{ \frac{\vert 216\left(B^2+C^2-1\right)\left(1-(B^2-C^2)^2\right)
-8\left(1+B^2+C^2\right)^3\vert}{1728\left(B^2+C^2-1\right)
\left(1-\left(B^2-C^2\right)^2\right)}}\right),&  \xi = -1,
\end{cases}
\end{equation}
where sign of $\xi$ is defined by the chosen solution (see~(\ref{19})
and Table~\ref{tab2}) of eqs.~(\ref{2}).

\section{Eigenvectors}
We previously used the Cardano method for calculating eigenvalues of the polynomial appearing
in (\ref{16}). We now face the problem of finding the eigenvectors. Let us consider
the eigenvector of this system
\begin{equation}
\label{36}
\begin{aligned}
A\Psi=\lambda\Psi
\end{aligned}
\Rightarrow
\begin{aligned}
\left(A-\lambda E\right)\Psi=0,
\end{aligned}
\end{equation}
where the matrix $A$ can be constructed from the system (\ref{15}) as
\begin{equation}
\label{37}
A=
\begin{pmatrix}
0          & -\delta_yC P_y &               \gamma_z\sqrt{1-P_z^2} \\
\gamma_xB\sqrt{1-P_x^2} &         0  &          -\delta_zP_z \\
-\delta_xBP_x &       \gamma_yC\sqrt{1-P_y^2} &           0
\end{pmatrix}.
\end{equation}
Vector $\Psi$ has the form
\begin{equation}
\label{38}
\Psi=
\begin{pmatrix}
U\\
V\\
W
\end{pmatrix},
\end{equation}
where $U$, $V$ and $W$ are components of the eigenvector.

We use the Gauss's method for the solution of (\ref{36}) to simplify
the factor $\left(A-\lambda E\right)$
\begin{equation}
\label{39}
\begin{gathered}
\left(A-\lambda E\right)=
\begin{pmatrix}
-\lambda          & -\delta_yC P_y &               \gamma_z\sqrt{1-P_z^2} \\
\gamma_xB\sqrt{1-P_x^2} &         -\lambda  &          -\delta_zP_z \\
-\delta_xBP_x &       \gamma_yC\sqrt{1-P_y^2} &           -\lambda
\end{pmatrix}
\sim\\
\sim
\begin{pmatrix}
-\lambda    & -\delta_yC P_y    &    &  \gamma_z\sqrt{1-P_z^2} \\
0 &   -\dfrac{\delta_y\gamma_xBCP_y\sqrt{1-P_x^2}}{\lambda}-\lambda  &  &
\dfrac{\gamma_z\gamma_xB\sqrt{1-P^2_z}\sqrt{1-P_x^2}}{\lambda}-\delta_zP_z \\
0 &  0 &   &   0
\end{pmatrix}.
\end{gathered}
\end{equation}
Matrix (\ref{39}) has one zero line and, therefore,  $U$ and $V$ can be defined via $W$,
and $W$ is free.

The system of linear equations is
\begin{equation}
\label{40}
\begin{aligned}
\begin{gathered}
-\lambda U  -\delta_yC P_y V + \gamma_z\sqrt{1-P_z^2}W=0, \\
\left(-\dfrac{\delta_y\gamma_xBCP_y\sqrt{1-P_x^2}}{\lambda}-\lambda\right)V  +
\left(\dfrac{\gamma_z\gamma_x B\sqrt{1-P^2_z}\sqrt{1-P_x^2}}{\lambda}-\delta_zP_z\right)W =0.
\end{gathered}
\end{aligned}
\end{equation}
and its solution is
\begin{equation}
\label{41}
\begin{gathered}
V  = \frac{
\gamma_z\gamma_xB\sqrt{1-P^2_z}\sqrt{1-P_x^2}-
\delta_z \lambda P_z}
{\delta_y\gamma_xBCP_y\sqrt{1-P_x^2}+\lambda^2}W,\\
U  =\frac{1}{\lambda} \left(-\delta_yC P_y \frac{
\gamma_z\gamma_xB\sqrt{1-P^2_z}\sqrt{1-P_x^2}-
\delta_z\lambda P_z}
{\delta_y\gamma_xBCP_y\sqrt{1-P_x^2}+\lambda^2} +
\gamma_z\sqrt{1-P_z^2}\right)W.
\end{gathered}
\end{equation}
If $\lambda\neq0$ we can let
\begin{equation}
\label{42}
\begin{aligned}
W  = \delta_y\gamma_xBCP_y\sqrt{1-P_x^2}+\lambda^2.
\end{aligned}
\end{equation}
Components $U$, $V$ and $W$ of the eigenvector can be computed as follows:
\begin{equation}
\label{43}
\begin{aligned}
&U=\frac{1}{\lambda}\left[-\delta_y C P_y \left(
\gamma_z\gamma_x B\sqrt{1-P^2_z}  \sqrt{1-P_x^2} - \delta_z \lambda  P_z \right)\right.+\\
&\qquad\left.\gamma_z\sqrt{1-P_z^2}\left(\delta_y \gamma_x BC P_y  \sqrt{1-P_x^2}+\lambda^2\right)\right],\\
&V=\gamma_z\gamma_xB\sqrt{1-P^2_z}\sqrt{1-P_x^2}-\delta_z\lambda P_z,\\
&W=\delta_y\gamma_xBCP_y\sqrt{1-P_x^2}+\lambda^2.
\end{aligned}
\end{equation}

\section{Conclusion}

Analytical expressions for coordinates of the stationary points and conditions for their
existence in the ABC flow are obtained. The coordinates are found from Eqs.~(\ref{8}).
The points exist in the region above the bifurcation line: $B^2+C^2=1$
(see Fig.~\ref{par1}). It has been proved that only 2 stationary points exist
at the line $B^2+C^2=1$. It is analytically shown that the stationary points
are of a saddle-node type (see Figs.~\ref{image1} and \ref{image2}) at all values of
the control parameters in the region of their existence. Exact expressions for the
eigenvalues (\ref{34}) and
eigenvectors (\ref{43}) of the stability matrix are given. The behavior of the stationary
points~(\ref{11}) along the bifurcation lines (\ref{10}) is considered.

\section*{Acknowledgments}
The authors would like to thank Prof. S. Prants for a critical reading the manuscript and valuable
comments. The work on stability and bifurcation analysis of the ABC flow
has been supported by the Russian Science Foundation
(project no.~16--17--10025) and
the work on finding and analysis of eigenvalues and eigenvectors has been supported by the Russian
Foundation for Basic Research (project no.~15--35--20105 mol-a-ved).

\bibliography{elsarticle-template-2-harv}
\bibliographystyle{model1-num-names}

\end{document}